\documentstyle[12pt]{article}

\normalbaselines
\begin{document}
\bibliographystyle{unsrt}
\pagenumbering {arabic}
\vbox{\vspace{38mm}}
\begin{center}
{\LARGE \bf NONLINEAR $W_{\infty}$ ALGEBRAS FROM NONLINEAR\\[2mm]
 INTEGRABLE DEFORMATIONS \\[2mm] OF SELF DUAL GRAVITY  }\\[5mm]

Carlos Castro \\{I.A.E.C 1407 Alegria\\
Austin, Texas 78757  USA }\\[3mm]
( May, 1994. Revised, Sept. 94)\\[5mm]
\end{center}
\begin{abstract}

A proposal for constructing a universal nonlinear ${\hat W}_{\infty}$ algebra
is
made as the symmetry algebra of
a rotational Killing-symmetry reduction of the
nonlinear perturbations of Moyal-Integrable
deformations of $D=4$ Self Dual Gravity (IDSDG). This is attained upon the
construction of a nonlinear bracket based on nonlinear gauge theories
associated with infinite dimensional Lie algebras.  A Quantization
and supersymmetrization program can also be carried out.
The relevance to the Kadomtsev-Petviashvili hierarchy, $2D$ dilaton gravity,
quantum gravity and black  hole physics is discussed in the concluding
remarks.

PACS : 0465.+e;0240.+m
\end{abstract}

\section{Introduction }

  Recently [1] a universal linear $W_{\infty}$ was constructed as the
symmetry algebra of a rotational Killing-symmetry reduction of the
Moyal integrable deformations of $D=4$ Self Dual Gravity (IDSDG) [2]. This
algebra turned out to be the $complexification$ of the direct sum of the
chiral and antichiral $W_{\infty},{\bar W}_{\infty} $ algebras. Central
extensions can be constructed by the cocycle formula in terms of the
logarithm of the derivative operator discussed in [3]. The latter work
was a complement of [4] where Fairlie and Nuyts using Moyal brackets
found a basis of differential operators on the circle which yield the structure
constants of the centerless linear $W_{\infty}$ [5]. An epimorphism map
(onto but not $1-1$ map) was established from the universal algebra into the
nonlinear ${\hat W}_{\infty}$ algebra of Wu and Yu [6] although the explicit
solution was not presented in [1]. A one parameter family of Hamiltonian
strutures for the Kadomtsev-Petviashvili (KP) hierarchy and a continous
deformation of nonlinear $W_{KP}$ algebras has been given by [7].
What is required now is a unifying geometrical framework to generate these
algebras.

The purpose of this letter is to construct nonlinear perturbations of the
$IDSDG$ based on a nonlinear bracket providing solutions to the epimorphisms
of [1]. Such nonlinear bracket is based on a nonlinear gauge theory principle
[8] which has attracted attention recently in connection to $2D$ dilaton
supergravity theories and Yang-Mills-like formulation of $R^2$ gravity with
dynamical torsion generalizing the Jackiw-Teitelboim's  model.
The nonlinear algebra (strictly speaking these are not Lie algebras) is based
on the complexification of the area-preserving diffs of a two-dim surface,
$sdiff~\Sigma$, which is essentially $SU^*(\infty)\sim SL(\infty , H)$
[9,10,20].
(quaternionic valued). Self Dual Yang-Mills equations (SDYM) based on this
group, after a suitable dimensional reduction and ansatz, lead to Plebanski's
second heavenly equation which furnishes solutions of SDG in $D=4,2+2$.
A Darboux change of variables converts the latter equation into the first
heavenly equation whose rotational Killing-symmetry reduction is linked to
$W_{\infty}$ algebras via the $SL(\infty)$ continous Toda equation [11].
Therefore, a nonlinear gauge theory based on $SU^*(\infty)$ SDYM in four
dimensions is linked to nonlinear ${\hat W}_{\infty}$ algebras [6].
It is in this fashion how we construct the new bracket which incorporates the
nonlinearities of [6].

The importance of Infinite Dimensional Lie Algebras and the Geometry of
Integrable Systems in connection to Self Dual Gravity has been emphasized
by many authors, in particular by Park [11], Strachan and Takasaki [2,12].
The supersymmetric extension was
carried out by the author [13]. Since the $sdiff~\Sigma$ Lie algebra
preserve the Poisson symplectic structure (the area in phase space)
an integrable deformation of these models was possible by means of the
Moyal bracket [14] while, at the same time, retaining the geometrical ideas
central to the integrability of SDYM/SDG.
The connection with the theory
of integrable systems came from the observation by Atiyah [15] that these
are reductions of SDYM systems in four or higher dimensions. For further
mathematical details pertaining the geometry of SDYM see Ward [16].

A non-linear extension
of these algebras and their quantization was provided by Wu and Yu [6] by
quantizing the
conformal non-compact coset model $sl(2,R)_k /U(1)$ in which the ${\hat
W}_{\infty}$ algebra appears as a hidden current algebra ( Bakas and
Kiritsis [18]).
Because this $sl(2,R)_k /U(1)$  coset model is connected to Witten's
black hole in $2D$ string theory, its quantization plays an important role
in understanding the physics of two-dim quantum gravity and black-holes [19]
.It is for this reason that a geometrical construction of a Universal nonlinear
${\hat W}_{\infty}$ algebra is warranted. This is the purpose of this work.

In $II$  [1] a candidate for a Universal linear $W_{\infty}$ algebra
is obtained from a Killing-symmetry reduction of Moyal-integrable deformations
of Self Dual Gravity (IDSDG) in $D=4$  based on the ideas by Strachan and
Takasaki [2,12].
In $III$ we present the nonlinear bracket extension of the Moyal algebra
and postulate the form of the equations governing the nonlinear perturbations
of
Moyal integrable deformations of SDG and the algebra of symmetry
transformations.
Finally, some brief remarks are made in relation to black-hole physics and
canonical quantum gravity. Complexified $D=4$ spacetime is
always assumed. $C^4$ is parametrized by the variables : $ y,z,{\tilde
y},{\tilde z}$. Moyal brackets are always assumed unless otherwise indicated.
Plebanski's first heavenly equation is $\{\Omega ,_y ,\Omega ,_z
\}_{{\tilde y},{\tilde z}} =1$;  where the bracket is the Poisson one.
Solutions of this equation yield complexified self dual metrics of the form
: $ds^2 =\Omega_{i {\tilde j}} dx^i d{\tilde x}^
j.~~x^i=y,z;~ and~{\tilde x}^j={\tilde y},{\tilde z}.$

\section{The Universal $W_{\infty}$ Algebra}

   We begin with by writing down the derivatives with respect to $y,{\tilde
y}$ (when acting on $\Omega$)
which appear in the Moyal integrable deformation of Plebanski`s first
heavenly equation to be discuss below. $r\equiv y {\tilde y}$.
 $$\partial_y
=(1/y)r\partial_r.~\partial_{\tilde y}
=(1/{\tilde y})r\partial_r.~\partial_y \partial_{\tilde y}=r\partial^2_r
+\partial_r.\eqno (1) $$
$$(\partial_{\tilde y})^2=(1/{\tilde y})^2 (r^2\partial_r^2 +r\partial_r).~~
(\partial_{\tilde y})^3 =(1/{\tilde y})^3 (r^3\partial^3_r
+r^2\partial^2_r -r\partial_r)......\eqno (2)$$

The Moyal bracket  is  obtained as a power expansion in the
deformation parameter $\kappa$ of the expression  :

$$\{f,g\}_{Moyal} \equiv [\kappa^{-1}sin~\kappa (\partial_{{\tilde
y}_f}\partial_{{\tilde z}_g}-\partial_{{\tilde
y}_g}\partial_{{\tilde z}_f})]fg.\eqno (3)$$

where the subscripts under ${\tilde y},{\tilde z}$ imply that the
derivatives act only on  $f$ or on $g$ accordingly.
In view of these  higher order derivatives in the Moyal bracket one can see
that
one is going to have a compatible power expansion ( of the solutions  to the
deformations  of Plebanski`s first heavenly equation [11,12]) with the
rotational Killing symmetry condition  : $\Omega (r\equiv y{\tilde
y};z,{\tilde z})$ if, and only if, one has the following power expansion for
$\Omega$ :
$$\Omega (y,{\tilde y},z,{\tilde z})\equiv
\sum_{n=0}^{\infty}~(\kappa/{\tilde y})^n \Omega_n (r,z,{\tilde z})\eqno (4)$$
where each $\Omega_n$ is a function $ only$ of $r,z,{\tilde z}$. Plugging
this expansion into the integrable deformation of Plebanski`s first
heavenly equation :
$$\{\Omega_z ,\Omega_y \}_{Moyal} =1.\eqno (5) $$
where the Moyal bracket is taken with respect to ${\tilde z},{\tilde y}$
yields an infinite family of differential equations with respect to the
variables $r,z,{\tilde z}$ only :

$$\{\Omega_{0z},\Omega_{0y}\}_{Poisson} =1.$$
$$0=\Omega_{0z{\tilde z}}[-\Omega_{1r}+(r\Omega_{1r})_r]
-r\Omega_{0rz}\Omega_{1r{\tilde z}}+$$
$$\Omega_{1z{\tilde z}}(r\Omega_{0r})_r +\Omega_{0r{\tilde z}}
(\Omega_{1z}-r\Omega_{1rz}). $$
$$..................................................\eqno (6)$$
where the subscripts represent partial derivatives of the functions
$\Omega_0,\Omega_1,.....$ with respect to the variables $r,z,{\tilde z}$ in
accordance with the
Killing symmetry reduction condition. The first equation, after a change
of variables, $is$ nothing but the $sl(\infty)$ continual Toda equation
[11]
whose symmetry algebra is the linear classical $w_{\infty}$ algebra [3].
The rest of the  equations are then the Moyal
integrable deformations of the continual Toda equation.

In order to determine the symmetry algebra of this infinite family we
follow closely
Strachan and Takasaki who  showed that an infinitesimal symmetry of
equation
(5) must be a solution of the  deformation of Laplace`s equation with respect
to a background-solution of (5) ( upon apllication of $\delta$ on (5)) :

$$\Delta_{\Omega} \delta \Omega \equiv \{\Omega_z,\delta\Omega_y
\}-\{\Omega_y,\delta\Omega_z\} =0. \eqno (7)$$
where $\Omega$ is a solution of (5).

A general straightforward solution of
(7) which was not given by [2,12], since these authors did not discuss
Killing symmetry reductions of (5), is  :

$$\delta \Omega \equiv \alpha [\epsilon L_{\Lambda}, \Omega] =
\{\Lambda({\tilde y}, {\tilde z}),\Omega
\}=\sum_{n=0}~(\kappa/{\tilde y})^n \delta \Omega_n (r,z,{\tilde z}). \eqno
(8)$$
due to the antisymmetry and Jacobi properties of the Moyal bracket and to
the fact that $\Omega$ is a solution of (5). $\alpha$ is a parameter whose
dimension is $(length)^2$ and it is needed to match dimensions. It will be
set to $unity$ in all of the equations that follow. $\Lambda$ is taken to
be dimensionless. $\epsilon$ is an infinitesimal parameter corresponding
to the infinitesimal symmetry and $L_{\Lambda}$ is a generator of the
universal $W_{\infty}$ algebra to be determined below. The generators of
the universal algebra are "parametrised" by a family of functions :
$$\Lambda ({\tilde y},{\tilde z}) =\sum_{n=0}
(\kappa/{\tilde y})^n {\tilde y}f_n ({\tilde z}). \eqno (9) $$

After performing the Moyal bracket in (8)  one has an infinite family of
equations
yielding the transformations $\delta \Omega_n (r,z,{\tilde z}),n=0,1,2....$ :

$$\delta \Omega_0 =r\Omega_{0r}f_{0{\tilde z}} -f_0\Omega_{0{\tilde z}}.$$
$$\delta \Omega_1 =- f_0\Omega_{1{\tilde z}}-f_{0{\tilde z}}\Omega_1 +
r\Omega_{1r}f_{0{\tilde z}}+r\Omega_{0r}f_{1{\tilde z}}   .$$
$$.................................  \eqno (10)$$

   The universal symmetry algebra is furnished as the  Lie algebra of
derivations on the space of solutions of (5):
$$[\delta_{\Lambda^1},\delta_{\Lambda^2}]\Omega =\delta_{\Lambda^3}
\Omega =\delta_{\Lambda^1 \otimes \Lambda^2}~\Omega. \eqno (11)$$

where $\Lambda^1 \otimes \Lambda^2 \equiv \{\Lambda^1,\Lambda^2\}$.
The Lie algebra of derivations is consistent with the Lie-Moyal bracket
structure present in the physics of Moyal integrable deformations of Self
Dual Gravity :

$$L_{\Lambda^1} \otimes L_{\Lambda^2}= [L_{\Lambda^1},L_{\Lambda^2}]
=L_{\Lambda^1 \otimes \Lambda^2} =L_{\{\Lambda^1,\Lambda^2 \}}. \eqno (12)$$

Given two functions $\Lambda^1 ({\tilde y},{\tilde z}),\Lambda^2 ({\tilde
y},{\tilde z})$ the Lie-Moyal structure yields $\Lambda^3$ in terms of the
former two functions after expanding in powers of $(\kappa/{\tilde y})^n$ :

$$ f^3_0 =f^2_0 f^1_{0{\tilde z}} -f^1_0 f^2_{0{\tilde z}}\not= 0.$$
$$f^3_1 =f^2_0 f^1_{1{\tilde z}} -f^1_0 f^2_{1{\tilde z}} \not= 0. $$
$$f^3_2 = (f^1_2 f^2_{0{\tilde z}} - 1\leftrightarrow 2)+(f^2_0
f^1_{2{\tilde z}} -1\leftrightarrow 2) \not=0.$$
$$...............\eqno (13)$$

Where we should keep in mind always the presence of the $\alpha$ parameter
which was set to $one$ in the above equations so that the dimensions match
properly. Therefore,
iteratively, we solve for the rotational Killing-symmetry
reduction of the Moyal-integrable deformations of Self Dual Gravity
(IDSDG) in $D=4$ and find its infinite dimensional universal $W_{\infty}$
algebra
which is the Moyal deformation of the $w_{\infty}$ algebra associated with
the $sl(\infty)$ continual Toda equation. Park [11] proved that this algebra
is in fact larger and was the $w_{\infty} \oplus {\bar w}_{\infty}$
(after a suitable real slice). Hence, the candidate universal linear
$W_{\infty}$
algebra is the $complexification$ of $W_{\infty} \oplus {\bar W}_{\infty}$.
Clearly this algebra is much bigger than $W_{\infty}$.

\section{ The Nonlinear $W_{\infty} $ Algebra}

We shall define the new bracket which incorporates the nonlinearities of
the $W$ algebras symbolically as  follows. Given two gauge potentials their
bracket is :
$$ [A_\mu ,A_\nu ] \equiv { \{A_\mu ,A_\alpha \} \over
{1 -\lambda F^{-1}_{\alpha \beta} \{A_\beta ,A_\nu \}}} . \eqno (14)$$
A power  expansion in $\lambda $ yields :
$$ [A_\mu,A_\nu ] \equiv \{A_\mu, A_\nu \} +\lambda \{A_\mu,A_\alpha \}
F_{\alpha \beta}^{-1} \{A_\beta,A_\nu \} +\lambda^2
\{A_\mu , \}F^{-1} \{ ,\} F^{-1}\{ ,A_\nu \} +..........\eqno (15).$$
$F_{\alpha \beta}^{-1}$ is the $inverse$ matrix function associated with the
field-strength of the $SU^*(\infty )$ Yang-Mills theory.
We recall that $SU^*(\infty)$ Yang-Mills theory [20]; i.e.the gauge
theory of the $sdiff~\Sigma$ group requires replacing Lie-algebra valued
gauge potentials by $c$-number $A_\mu (x^m;q,p)$ objects depending on the two
extra internal variables parametrising the surface $\Sigma$
in addition to $x^m$. Lie-algebra commutators are replaced by Poisson brackets.

The bracket (15) is reminiscent of Dirac's bracket in the quantization of gauge
systems with constraints. The bracket $linearizes$ in the special case that the
gauge potentials are spacetime $independent$ :
$$F_{\mu \nu} \equiv \partial_\nu A_\mu - (\mu \leftrightarrow \nu ) +
\{A_\mu, A_\nu \}\Rightarrow \{A_\mu, A_\nu \}. \eqno (16) $$
The bracket is the Moyal bracket with respect to the internal variables
if we wish to make contact with
the nonlinear ${\hat W}_{\infty}$ algebra of Wu and Yu [6].
Summation over repeated indices is implied.
The bracket's antisymmetry and derivation properties are trivially satisfied
by inspection.$[A_\mu,A_\nu ]=-[A_\nu,A_\mu]$ and $[A,BC]=[A,B]C+B[A,C]$.
This is due to the bracket nature of the last Moyal bracket in each one of
the terms in the r.h.s of (15).
The Jacobi property  is more subtle and can be verified if one recurs to
the vector calculus result :
$$\vec {A} \wedge (\vec {B} \wedge \vec {C}) +cyclic =0. \eqno (17)$$
Per example,  the term :$ \{A,M \} F^{-1}_{MN} \{N ,B \}$ ca be represented
symbolically, after a matrix multiplication of three antisymmetric matrices,
as  :
$$
(\vec {A} \wedge \vec {B }) \phi(\vec {M},\vec {N}).
 \eqno (18)  $$
$\phi$ is a scalar valued function of $\vec {M},\vec {N}$. One learns that
$\phi (\vec {M},\vec {N}) =\phi (\vec {N},\vec {M})$.
This is due to the fact that the bracket $[ , ]$ is a map from
$\cal A \otimes \cal A \rightarrow \cal A$. The vector space of the gauge
potentials is represented by $\cal A$. This explains
why one must contract vector indices as shown in eq-(18).
It is for
this reason that one must have insertions of the $inverse$ matrix function
between two Moyal brackets exactly as it occurs in Dirac's prescription for
quantization of constrained Hamiltonian systems.
A structure similar to (14,15) appeared in the free field realizations of
${\hat W}_{\infty}$ given in [6,18] where an explicit form of the $KP$ first
order pseudo-differential operator was given as ( $D=d/dz$) :
$$L=D+\sum^{\infty}_{r=0}~u_rD^{-r-1} =D+{\bar j}{1\over {D-({\bar j}
+j)}}j. \eqno (19)$$
Therefore the bracket $[A,[B,C]]$  can be represented symbolically
as a sum of terms of the form :
$$\vec {A}\wedge (\vec {B}\wedge \vec {C}) [1+\phi +\phi^2 +....]^2
\eqno (20) $$
which upon taking the cyclic sum one gets zero.
Since the ${\hat W}_{\infty}$
algebras are arbitrarily nonlinear ( in powers of $\lambda$) one must sum over
an arbitrary number of terms. Setting $\lambda =0$ one recovers the Moyal
algebra which was shown to be isomorphic to the centerless linear $W_{\infty}$
algebra in [3,4]. Because the symmetry algebra is the $complexification$ of the
direct sum of a chiral and antichiral $W_{\infty}, {\bar W}_{\infty}$ algebra,
respectively [1], in order to establish the isomorphism, one first
has to take a suitable real slice and then project into the chiral sector.
The symmetry algebra of the nonlinear perturbations of the Moyal integrable
deformations of Self Dual Gravity ( after a Killing symmetry reduction )
is larger than the ${\hat W}_\infty$
algebra. For this reason we coined them "universal" $W_{\infty}$ algebras [1].
Central terms can be incorporated by adding cocycles to the algebras and these
can be
expressed in terms of the logarithm of the derivative operator given in [3].
Cocycles in terms of the symmetries of the $tau$ function were also given
previously by Takasaki [12].
If one had used Poisson brackets, instead, one would obtain nonlinear
deformations of the Bakas algebra $w_{\infty}$. A Quantization program
can be carried out by performing a quantum deformation of the algebra as it was
performed in [6] to obtain the Quantum ${\hat W}_{\infty}$ algebra.

Having written down the new bracket we can conjecture that the nonlinear
perturbations of Moyal integrable deformations of Self
Dual Gravity can be written in terms of the Plebanski 's second
heavenly form $\Theta (y,\tilde y,z, \tilde z)$ as follows [9,10].  Start with
the complexified  $SU^* (\infty)$ Self Dual Yang-Mills equations in
complexified spacetime, $C^4$, of signature $(4,0),(2,2)$, respectively ,
$F_{y\tilde y} +(-)F_{z\tilde z} =0$ and $F_{yz} =F_{\tilde y \tilde z}
=0$. The curvatures must be computed using the nonlinear bracket (14,15).
We learnt from [9,10] that, in the case of ordinary Poisson brackets, a
dimensional reduction : $\partial_y =\partial_{\hat
 q}$; $-\partial_{\tilde y} =\partial _{\hat p}$ and the ansatz of [9,10],
where the hatted variables $\hat q, \hat p$ represent the complexification of
the $sdiff~\Sigma$, yields Plebanski's second heavenly equation [9,10] :

$$(\Theta,_{\hat p \hat q})^2 -\Theta,_{\hat p \hat p} \Theta,_{\hat q
\hat q} +\Theta,_{z\hat q} -\Theta,_{\tilde z \hat p} =0. \eqno (21)$$

The semicolon means partial derivatives with respect to the corresponding
variables. Eq-(21) yields self-dual solutions to the complexified Einstein's
equations providing a family of hyper-Kahler metrics on the complexification
of the cotangent space of $\Sigma$. $\Theta$ represents a continuous self dual
deformation of the flat metric in $T^*\Sigma$. The field-strengths are now
computed with the nonlinear bracket given by (15) so that
(21) is modified accordingly. The symmetry algebra of the space of
solutions of (21) originates from the original YM gauge invariance of the
$SU(\infty)$ SDYM equations :
$$ \delta A_\mu = D_{A_\mu}\epsilon.~\delta
F_{\mu\nu}=[F_{\mu\nu},\epsilon].\eqno (22)$$

The YM potentials, as functions of $\Theta$, were given in [9,10] :
$A_y =\Theta,_y.~A_{\tilde y}=\Theta,_{\tilde y}....$. Therefore,
symmetries under the transformations (22) determine those of $\Theta$. Due
to the nonlinearities of (15) the field strength does not transform
homogeneously unless the gauge parameter is a suitable judicious field
dependent quantity, $\epsilon (A)$ obeying an equation like the one (26)
below.

Where do the $W$ symmetries appear ? From the underlying KP equation . A
detailed construction was given
in [9] mapping solutions of the KP equation into the dimensionally-reduced
solutions (from $4$ to $3$) of the $D=2+2$ Plebanski's second heavenly equation
after a  certain
tuning (constraints) between the YM potentials was imposed. The
dimensionally-reduced
Plebanski equation turned  out to be after one showed that $F_{y {\tilde y}}=0$
:
$$F_{z{\tilde z}}= \Theta,_{zq}-\Theta,_{{\tilde z}p}=0.\eqno(23)$$

 We will see shortly the importance of (23). When (15) is implemented one
has a deformation of (23) :${\cal F}_{z {\bar z}} =0$. The latter implies
that (23) receives $\lambda$ corrections so that $F_{z {\bar z}}$ is no
longer zero. A zero value for $F$ will also render eq-(15) ill-defined : it
will be singular.

Boer and Goeree in a series of
articles [21] constructed a covariant $W$ gravity action and provided with
a geometrical interpretation of $W$ transformations ( a $W$ algebra) from
Gauge Theory for those $W$ algebras related to embeddings of $sl(2)$ in a
Lie algebra ${\bf g}$. The $W$ transformations were just homotopy
contractions of ordinary gauge transformations. Their starting point was
to constrain the $A_z$ potential to have the form :$A_z ={\bf \Lambda} +{\bf
W}$
where $\Lambda$ is a constant matrix and the components of ${\bf W}$
transform as fields of certain spins determined by a gradation of ${\bf g}$:
$${\bf W}=\sum~W^i (z,{\bar z})X_i.~{\bf g}=\oplus g_{\alpha_i}.\eqno (24)$$
$X_i$ are the generators within each member $g_{\alpha_i}$ of the gradation
and $W^i$ are the
generators of the $W$ algebra. So we can see how the YM potentials contain
the $W$ generators (currents).

  The second step was to
find those gauge transformations that leave the form of $A_z$ invariant :
$$\delta A_z =D_{A_z}(A_{\bar z}).~A_{\bar z}={1\over{1+\lambda L(\partial +
Ad_W)}}Kernel(Ad_\Lambda). \eqno (25)$$

$\lambda$ is a parameter which will be identified with the one in (15). $L$ is
the homotopy (``integration'') operator which roughly maps Lie-algebra-valued
$p$
forms in $R^n$ to $(p-1)$ forms on $R^{n-1}$. It is defined as the inverse
of the adjoint action of $\Lambda: Ad_\Lambda =[\Lambda, *]$. The
expression relating $A_{\bar z}$ in terms of the sum of a series of operators
acting on the kernel of the $Ad_\Lambda$; i.e. in terms of  $A_z$, was obtained
from the
zero curvature condition :$F(A_z, A_{\bar z}) =0$. Now we can see the
relevance of eq-(23). Finally, what is required  is to find the form of the
$field-dependent$ gauge parameter :$\epsilon (A)$ appearing in (22) in order
to match the $W$ transformations of (25) in the $N\rightarrow \infty$
limit. We shall refer now to $sl(\infty, H)\sim su^*(\infty)$. One has
:${\bf \Lambda}\rightarrow \Lambda (q,p).{\bf W}\rightarrow W(z, {\bar
z},q,p)$....

Writing the expression for $A_{\bar z}$ as $A^o_{\bar z}+\sum~\lambda^n
A^{(n)}_{\bar z}$;  and a similar expression for $\epsilon
(A)=\sum~\lambda^n \epsilon^{(n)}(A)$
allows to determine, order by order in $\lambda$, the required form of each
of the terms appearing in the expansion of $\epsilon (A)$ :

$$\partial_z \epsilon^o +\{A_z,\epsilon^o \}=(\partial_z +\{A_z,*\})
A^o_{\bar z}.$$

$$\partial_z \epsilon^1+ \{A_z, \epsilon^1 \}
+\{A_z,A_\alpha\}F^{-1}_{\alpha\beta}\{A_\beta,\epsilon^o \}=(\partial_z
+\{A_z,*\})A^1_{\bar z}. $$
$$...............................\eqno (26)$$

The YM potentials appearing in the l.h.s of (26) are solutions to :${\cal
F}_{z {\bar z}}=0$ and should not be confused with the solutions to the
$F_{z {\bar z}} =0$ . Eq-(26) will ensure that ${\cal F}$ transforms as it
should since (26) is part of
$$\delta_W F_{z {\bar z}}=\delta {\cal F}_{z {\bar z}} =[{\cal F}_{z
{\bar z}},\epsilon (A)]=0.\eqno (26b)$$

due to the fact that ${\cal F}_{z {\bar z}} =0$. For consistency one ought
to add a holomorphic function $f(z,\lambda)$ to the r.h.s of (26).
The brackets appearing in (26) are Poisson brackets. Upon quantization one
replaces them by the Moyal brackets. This we learnt from [5] that a
quantization of $w_{\infty}$ deforms into the linear $W_{\infty}$. The
brackets are computed from the relationship :
$$ \{f(A_i),g(A_j)\} = {\delta f\over {\delta A_i}}\{A_i,A_j\}{\delta
g\over {\delta A_j}}. \eqno (27)$$

Given two solutions $\epsilon_1, \epsilon_2$ to the functional differential
equation given by (26) one has, by construction, that  these $W$
transformations satisfy a nonlinear $W_{\infty}$ algebra.
$$[\delta_{\epsilon_1}, \delta_{\epsilon_2}] {\cal F}=\delta_{\epsilon_3}
{\cal F}=0.~\epsilon_3 =[\epsilon_1,\epsilon_2]. \eqno (28)$$
This is due to the Jacobi property of the bracket (15).
If one wishes to recast the above in terms of Plebanski's first heavenly
equation one can. A Darboux change of variables transforms the former
equation (21)  into Plebanski's
first heavenly equation. When a Moyal bracket is used we learnt [2] that
the latter equation deforms into $\{\Omega_z, \Omega_y\}=1$.
When the nonlinear bracket is being used, instead, such equation should
represent the nonlinear perturbations associated with the Moyal deformations
of Self Dual Gravity. This equation reads
$$[\Omega_z (A_\mu),\Omega_y (A_\nu) ] =1.  \eqno (29)$$

In order to compute this last bracket in terms of (15)  one must insert the
dependence of $\Omega$ on the potentials due
to the  Darboux change of coordintes which is essentially a Backlund-type of
transformation relating derivatives of $\Theta$ to those of $\Omega$. Since the
YM
potentials are essentially derivatives of $\Theta$
the connection may be established in order to evaluate (28). One must also
include integration
``constants'' (functions) which are not purely arbitrary because one has to
satisfy (28) as well as the nonlinear generalization of (21) which
constrains the possible solutions of the YM potentials.
A further rotational Killing symmetry reduction of (28) in the lines of $II$
yields an
infinite family of equations whose symmetry algebra is the $complexification$
of ${\hat W}_{\infty}\oplus {\overline {\hat W}}_{\infty}$. This is the main
result of this letter completing the results in [1].
To conclude : Nonlinear Gauge Theories [8] of the $SU^*(\infty)$ SDYM theory
$contain$ (after dimensional reductions and/or
Killing symmetry reductions)
the nonlinear ${\hat W}_{\infty}$ algebras.
A quantization of the Killing symmetry reductions of nonlinear
perturbations of the $IDSDG$ in $D=4$ can be
achieved thanks to the Quantum
${\hat W}_{\infty}$ algebra constructed recently by Wu and Yu [6].
This quantum algebra contains all algebras in the classical limit after
apropriate truncations and contractions. It was seen as
a realization of the hidden symmetries of the quantized non-compact coset
model $sl (2,R)_k/U(1)$ [18].

Perhaps the most important consequence of the quantization
based on the $sl(2,R)_k/U(1)$ coset
model is the connection of this coset model to the black hole in $2D$
string theory found by Witten [19]. The construction of the Quantum
version of the $KP$ hierachy provided for us an infinite
set of commuting quantum charges in an explicit and closed form.
It is not difficult to see the importance that these infinite number of
conserved Quantum charges will have for the Quantum ${\hat W}_{\infty}~Hairs$
of
Witten's $2D$ stringy-black hole solution associated with the
gauged-$sl(2,R)_k/U(1)$ WZNW models.
What would be the situation in the full-fledged $4D$ Gravitational theory
versus the Killing-symmetry reductions of nonlinearly-perturbed $IDSDG$
(an effective $3D$
theory )? The work of Ashtekar and others [22] on the loop representation of
canonical quantum gravity overlaps with ours in the sense that we had
started with
a complexified self dual $SU^* (\infty)$ Yang-Mills theory which led to Self
Dual complexified Gravity upon dimensional reduction [9,10];
and  the effective
$3D$ theory upon the
Killing-symmetry reduction should bear a connection to the Knot/Chern-Simmons
theoretical formulations of $3D$ gravitational theory.
The fact that it
is $SU^*(\infty)$ SDYM
theory the one that generates the lower dimensional integrable models and their
hierarchies points that its moduli space must contain important
information pertaing the Geometry of Strings, IDSDG, black holes,
etc....To conclude : IDSDG +nonlinear-integrable
perturbations+Quantization $\sim$ Quantum-Nonlinear $W_{\infty}$ algebra.

{\bf Acknowledgements}.
We thank J.A. Boedo, J. Rapp and T.Grewe for their hospitality at the KFA
in Julich, Germany; and to Tanya Stark, Anke Kaltenbach and Wolfe Tode
for providing a warm and stimulating
atmosphere to carry out this work.

\section {REFERENCES}

1. C. Castro : " A Universal $W_{\infty}$ algebra and Quantization of

Integrable Deformations of Self Dual Gravity " I.A.E.C-2-1994 preprint.

2. I.Strachan : Phys.Letters {\bf B 282} (1992) 63.

3. I. Bakas : Comm. Math.  Phys. {\bf 134} (1990) 487.

I.Bakas, B.Khesin and E. Kiritsis : Comm. Math. Phys. {\bf 151} (1993) 233.

4. D.B. Fairlie and J. Nuyts : Comm. Math. Phys. {\bf 134} (1990) 413.

5. C. Pope,  L. Romans and X.  Shen :Phys.   Letters {\bf B 236} (1990) 173.

6. F. Yu and Y.S. Wu ; Jour.  Math.  Phys.  {\bf 34} (1993) 5851. ibid

{\bf 34} 5872. Nucl.Phys.{\bf B 373} (1992) 713.

7.J.Figueroa-O'Farrill, J. Mas and E. Ramos : preprint BONN-HE-92-20.

8. N. Ikeda : "$2D$ Gravity and Nonlinear Gauge Theory" Kyoto-RIMS-953-93.

9.  C. Castro :  Jour.Math.Phy. {\bf 35} no. 6 (1994) 3013.

10. C. Castro : Jour. Math. Phys. {\bf 34} (1993) 681.

11. Q.H. Park : Int. Jour. Mod.  Phys. {\bf A7} (1991) 1415.

12. K. Takasaki and T. Takebe : Lett. Math. Phys. {\bf 23} (1991) 205.

13. C.Castro . Journal of Math.   Phys. {\bf 35 no.2} (1994) 920.

14.  J. Moyal : Proc. Cam. Phil.  Soc. {\bf 45} (1945) 99.

15. M. Atiyah : in $Advances~in~Mathematics~Supplementary~Studies~vol.7A$.

Academic Press, New York, 1981.

16. R.S. Ward : Phys. Letters {\bf A 61} (1977) 81.

17. M.J. Ablowitz and P.A. Clarkson :{\bf "Solitons, Nonlinear Evolution

Equations ans Inverse Scattering "}.  London Math. Soc. Lect.   Notes {\bf

149}; Cambridge University Press 1991. Comm.Math.Phys. {\bf 158} (1993) 289.

18. I. Bakas and E. Kiritsis : Int. Journal of Mod.Phys.{\bf A7} (1992) 55.

19. E.  Witten :  Phys.Rev. {\bf D 44} (1991) 314.

20. E. G. Floratos, J. Iliopoulos, G. Tiktopoulos : Phys. Let. {\bf B 217}

(1989) 285.

21. J. de Boer and J. Goeree :''Covariant $W$ Gravity and Its Moduli Space
from Gauge Theory'' . Utrecht University THU-92-14 preprint. July 92.
Phys. Lett. {\bf 274 B} (1992) 289.

22. A. Ashtekar and R. Tate : " Lectures on non-perturbative quantum gravity"

World Scientific, Singapore (1991) and references therein.

\end{document}